\RequirePackage{fixltx2e}
\documentclass[a4paper]{isp}

\usepackage[american]{babel}
\usepackage{graphicx}

\usepackage{paralist}

\usepackage[T1]{fontenc}

\usepackage{lmodern}

\usepackage{microtype}
\usepackage{amsmath, amsthm}
\usepackage{booktabs}
\usepackage{url}

\ifnum\pdfoutput>0
\usepackage[
bookmarks=false,
breaklinks=true,
colorlinks=true,
linkcolor=black,
citecolor=black,
urlcolor=black,
pdfpagelayout=SinglePage
]{hyperref}
\usepackage[all]{hypcap}
\else
\usepackage{hyperref}
\fi

\usepackage[capitalise,nameinlink]{cleveref}
\crefname{section}{Sect.}{Sect.}
\Crefname{section}{Section}{Sections}
\crefname{figure}{Fig.}{Fig.}
\Crefname{figure}{Figure}{Figures}

\usepackage{xspace}

\def\aap{A\&A\,  }
\def\apj{ApJ\,  }
\def\apjs{ApJS  }
\def\araa{ARA\&A  }

\def\mnras{MNRAS\,  }

\def\pasp{PASP  }

\def\2F1{~_2F_1}
\def\sun{\hbox{$\odot$}}
\def\sunmass{\mathcal{M}_{\sun}}

\begin{document}
\pdfgentounicode=1
\title
{
A left and right  truncated lognormal distribution for the stars
}

\author{L. Zaninetti}
\institute{
Physics Department,
 via P.Giuria 1, I-10125 Turin,Italy \\
 \email{zaninetti@ph.unito.it}
}

\maketitle
\begin {abstract}
The initial mass function for the stars
is often modeled by a lognormal
distribution.
This paper is devoted  to demonstrating  the advantage
of introducing  a left and right   truncated lognormal
probability
density function, which is characterized
by four parameters.
Its normalization constant,
mean, the variance, second moment about the origin
and distribution function are calculated.
The chi-square  test and the Kolmogorov--Smirnov test are  performed
on four  samples  of stars.
\keywords
{
Stars: characteristics and properties of
Stars: normal
}
\end{abstract}

\section{Introduction}

The initial  mass
function (IMF) for the stars 
was {\it firstly} 
fitted with a power law by Salpeter, see  \cite{Salpeter1955}.
He suggested
$p ( {{m}}) \propto   {{m}}^{-\alpha}$
where $p   ( {{m}})$ represents  the probability
of having  a mass between $ {{m}}$ and
$ {{m}}+d{{m}}$  and 
he found  $\alpha= 2.35$
in the range $10  {M}_{\sun}~>~ {M} \geq  1  {M}_{\sun}$.
{\it Secondly} the IMF was fitted  with three  power laws, see
\cite{Scalo1986,Kroupa1993,Binney1998} and four power laws, see
\cite{Kroupa2001,Bastian2010,Kroupa2013}. 
The piecewise broken inverse power law IMF 
is 
\begin{equation}
p(m) \propto m^{-\alpha_i} \quad,
\end {equation}
each zone  being  characterized  by a different
exponent  ${\alpha_i}$
and two boundaries $m_i$ and $m_{i+1}$. 
In order to have a probability density function (PDF) 
normalized  to unity, one must have
\begin{equation}
\sum _{i=1,n}  \int_{m_i}^{m_{i+1}} c_i m^{-\alpha_i} dm =1
\quad.
\label{uno}
\end{equation}
The number of parameters to be found from the considered 
sample for the $n$-piecewise IMF is $2n-1$ when
$m_1$ and $m_{n+1}$ are the minimum  and maximum  of the masses 
of the sample.
In the case of $n=4$, which fits also the 
region of brown dwarfs (BD), see
\cite{Zaninetti2013a},
the number of  parameters is seven.
In the field of statistical distributions, 
the PDF  
is usually defined by two parameters.
Examples of two-parameter PDFs are:
the beta, gamma, normal, and lognormal  distributions,
see \cite{evans}.
The lognormal distribution is widely used
in order to model the IMF  for the
stars, see \cite{Larson1973,Miller1979,Zinnecker1984,Chabrier2003}.
The lognormal distribution is defined in the range
of  $\mathcal{M}\, \in (0, \infty)$  where $\mathcal{M}$
is the mass of the star.
Nevertheless, the stars have minimum and maximum  values,
as an example  from the MAIN SEQUENCE,  an M8 star has
$\mathcal{M} = 0.06 \sunmass$ and an O3 star has
$\mathcal{M} = 120  \sunmass$, see \cite{Cox}.
The presence  of boundaries for the stars
makes attractive the analysis of a left and right
truncated  lognormal.
In Section \ref{lognormal}, the structure of the lognormal
distribution is reviewed.
In Section  \ref{lognormaltrunc}, the truncated lognormal
distribution is derived.
In Section \ref{applications}, a  comparison between
the lognormal and truncated lognormal is done on four
catalogs of stars.
In Section \ref{others}, we compare the results of the truncated
lognormal distribution with the
double Pareto lognormal,
the truncated beta, and
the truncated gamma distributions.

\section{The lognormal distribution}
\label{lognormal}

Let $X$ be a random variable
defined in
$[0, \infty]$;
the {\em lognormal}
PDF, 
following \cite{evans}
or formula (14.2)$^\prime$ in
\cite{univariate1}, is
\begin{equation}
PDF (x;m,\sigma) = \frac
{
{{\rm e}^{-\,{\frac {1}{{2\,\sigma}^{2}} \left( \ln  \left( {
\frac {x}{m}} \right )  \right ) ^{2}}}}
}
{
x\sigma\,\sqrt {2\,\pi}
}
\quad,
\label{pdflognormal}
\end{equation}
where $m$ is the median and $\sigma$ the shape parameter.
The distribution function (DF) is
\begin{equation}
DF (x;m,\sigma) =
\frac{1}{2}+\frac{1}{2}\,{\rm erf} \left(\frac{1}{2}\,{\frac {\sqrt {2} \left( -\ln  \left( m
 \right ) +\ln  \left( x \right )  \right ) }{\sigma}}\right )
\quad ,
\end{equation}
where ${\rm erf(x)}$ is the error function, defined as
\begin{equation}
\mathop{\mathrm{erf}\/}\nolimits
(x)=\frac{2}{\sqrt{\pi}}\int_{0}^{x}e^{-t^{2}}dt
\quad ,
\end{equation}
see \cite{NIST2010}.
The average  value or mean, $E(X)$,  is
\begin{equation}
E (X;m,\sigma) = m{{\rm e}^{\frac{1}{2}\,{\sigma}^{2}}}
\quad ,
\label{xmlognormal}
\end{equation}
the variance, $Var(X)$, is
\begin{equation}
Var=
{{\rm e}^{{\sigma}^{2}}} \left({{\rm e}^{{\sigma}^{2}}}-1 \right ) {m
}^{2}
\quad,
\label{varlognormal}
\end{equation}
the second moment about the origin, $E^2(X)$, is
\begin{equation}
E (X^2;m,\sigma) = {m}^{2}{{\rm e}^{2\,{\sigma}^{2}}}
\quad .
\label{momento2lognormal}
\end{equation}
The experimental sample  consists of the data $x_i$ with
$i$  varying between  1 and  $n$;
the sample mean, $\bar{x}$,
is
\begin{equation}
\bar{x} =\frac{1}{n} \sum_{i=1}^{n} x_i
\quad ,
\label{xmsample}
\end{equation}
the unbiased sample variance, $s^2$, is
\begin{equation}
s^2 = \frac{1}{n-1}  \sum_{i=1}^{n} (x_i - \bar{x})^2
\quad ,
\label{variancesample}
\end{equation}
and the sample $r$th moment  about the origin, $\bar{x}_r$,
is
\begin{equation}
\bar{x}_r = \frac{1}{n} \sum_{i=1}^{n} (x_i)^r
\quad .
\label{rmoment}
\end{equation}
The parameter estimation is here obtained in two ways.
The matching moments estimator, (MME), is the first method:
\begin{equation}
E (X;m,\sigma)  = \bar{x}_1 \quad ;  \quad 
E (X^2;m,\sigma)= \bar{x}_2  \quad ,
\end{equation}
and therefore
\begin{eqnarray}
\widehat{m}      = {\frac {{\bar{x}_{{1}}}^{2}}{\sqrt {\bar{x}_{{2}}}}}      
\nonumber \\
\widehat{\sigma} = \sqrt {2}\sqrt {\ln  \left( {\frac {\sqrt {\bar{x}_{{2}}}}{\bar{x}_{{1}}}}
 \right ) }
 \quad .
\end{eqnarray}
The {\it second} method implements the maximum-likelihood estimation 
(MLE), see \cite{evans}.

\section{The truncated lognormal distribution}

\label{lognormaltrunc}

Let $X$ be a random variable
defined in
$[x_l, x_u ]$;
the {\em truncated lognormal}
PDF    ($PDF_T$) is
\begin{eqnarray}
PDF_T (x;m,\sigma,x_l,x_u) = \\
\frac{
\sqrt {2}{{\rm e}^{-\frac{1}{2}\,{\frac {1}{{\sigma}^{2}} \left( \ln  \left( {
\frac {x}{m}} \right )  \right ) ^{2}}}}
}
{
-\sqrt {\pi}\sigma\, \left( {\rm erf} \left(\frac{1}{2}\,{\frac {\sqrt {2}}{
\sigma}\ln  \left( {\frac {x_{{l}}}{m}} \right ) }\right )-{\rm erf}
\left(\frac{1}{2}\,{\frac {\sqrt {2}}{\sigma}\ln  \left( {\frac {x_{{u}}}{m}}
 \right ) }\right ) \right ) x
 }
\quad,
\label{pdflognormaltruncated}
\end{eqnarray}
where $m$ is now the  scale parameter,
$\sigma$ is the shape parameter,
$x_l$  denotes the minimal value, and
$x_u$  denotes the maximal value.
The introduction of the following coefficients
allows a compact notation
\begin{equation}
a_1 = \frac{1}{2}\,{\frac {\sqrt {2} \left( -{\sigma}^{2}+\ln  \left( x_{{l}}
 \right ) -\ln  \left( m \right )  \right ) }{\sigma}}
 \quad ,
 \nonumber
\end{equation}
 \begin{equation}
a_2 = \frac{1}{2}\,{\frac {\sqrt {2} \left( {\sigma}^{2}+\ln  \left( m \right ) -\ln
 \left( x_{{u}} \right )  \right ) }{\sigma}}
 \quad ,
 \nonumber
\end{equation}
\begin{equation}
a_3 =\frac{1}{2}\,{\frac {\sqrt {2} \left( \ln  \left( x_{{l}} \right ) -\ln
 \left( m \right )  \right ) }{\sigma}}
 \quad ,
 \nonumber
\end{equation}
\begin{equation}
a_4 =\frac{1}{2}\,{\frac {\sqrt {2} \left( -\ln  \left( x_{{u}} \right ) +\ln
 \left( m \right )  \right ) }{\sigma}}
 \quad ,
 \nonumber
\end{equation}
\begin{equation}
a_5 =\frac{1}{2}\,{\frac {\sqrt {2} \left( -2\,{\sigma}^{2}+\ln  \left( x_{{l}}
 \right ) -\ln  \left( m \right )  \right ) }{\sigma}}
 \quad ,
 \nonumber
\end{equation}
 \begin{equation}
a_6 =\frac{1}{2}\,{\frac {\sqrt {2} \left( 2\,{\sigma}^{2}+\ln  \left( m \right ) -
\ln  \left( x_{{u}} \right )  \right ) }{\sigma}}
 \quad ,
 \nonumber
\end{equation}
\begin{equation}
a_7 =\frac{1}{2}\,{\frac {\sqrt {2} \left( -2\,{\sigma}^{2}+\ln  \left( x_{{u}}
 \right ) -\ln  \left( m \right )  \right ) }{\sigma}}
 \quad ,
 \nonumber
\end{equation}
\begin{equation}
a_8 = \frac{1}{2}\,{\frac {\sqrt {2} \left( \ln  \left( x_{{u}} \right ) -\ln
 \left( m \right )  \right ) }{\sigma}}
 \quad .
 \nonumber
\end{equation}
In the compact notation the PDF is
\begin{equation}
PDF_T (x;m,\sigma,x_l,x_u) = \frac
{
-\sqrt {2}{{\rm e}^{-\frac{1}{2}\,{\frac {1}{{\sigma}^{2}} \left( \ln  \left(
{\frac {x}{m}} \right )  \right ) ^{2}}}}
}
{
\sqrt {\pi}\sigma\, \left( {\rm erf} \left(a_{{3}}\right )-{\rm erf}
\left(a_{{8}}\right ) \right ) x
}
\quad,
\label{pdflognormaltruncatedcompact}
\end{equation}
the DF  is
\begin{equation}
DF_T (x;m,\sigma,x_l,x_u)=
\frac
{
-{\rm erf} \left(\frac{1}{2}\,{\frac {\sqrt {2}}{\sigma}\ln  \left( {\frac {x
}{m}} \right ) }\right )+{\rm erf} \left(a_{{3}}\right )
}
{
{\rm erf} \left(a_{{3}}\right )-{\rm erf} \left(a_{{8}}\right )
}
\quad ,
\end{equation}
the mean, $E(X)_T$,  is
\begin{equation}
E_T (X;m,\sigma,x_l,x_u) =
\frac
{
{{\rm e}^{\frac{1}{2}\,{\sigma}^{2}}}m \left( {\rm erf} \left(a_{{1}}\right )+
{\rm erf} \left(a_{{2}}\right ) \right )
}
{
{\rm erf} \left(a_{{3}}\right )+{\rm erf} \left(a_{{4}}\right )
}
\quad ,
\label{xmlognormaltruncated}
\end{equation}
the variance, $Var_T(X)$, is
\begin{equation}
Var_T(X;m,\sigma,x_l,x_u) =
\frac
{
N
}
{
\left( {\rm erf} \left(a_{{3}}\right )+{\rm erf} \left(a_{{4}}\right )
 \right ) ^{2}
}
\quad ,
\end{equation}
where
\begin{eqnarray}
N={{\rm e}^{{\sigma}^{2}}}    \bigg( {\rm erf} (a_{{3}}   )
{\rm erf}    (a_{{5}}   ){{\rm e}^{{\sigma}^{2}}}+{\rm erf}
   (a_{{3}}   ){\rm erf}    (a_{{6}}   ){{\rm e}^{{\sigma}^{
2}}}+{\rm erf}    (a_{{4}}   ){\rm erf} (a_{{5}}   ){
{\rm e}^{{\sigma}^{2}}}
\\
+{\rm erf}    (a_{{4}}   ){\rm erf} (
a_{{6}}   ){{\rm e}^{{\sigma}^{2}}}-    ( {\rm erf} (a_{{1}}
   )    ) ^{2}-2\,{\rm erf}    (a_{{1}}   ){\rm erf} (
a_{{2}}   )-    ( {\rm erf}    (a_{{2}}   )    ) ^{2}
    \bigg) {m}^{2}
\quad ,
\end{eqnarray}
the second moment about the origin, $E_T^2(X)$, is
\begin{equation}
E_T (X^2;m,\sigma,x_l,x_u) =
\frac
{
-{{\rm e}^{2\,{\sigma}^{2}}}{m}^{2} \left( -{\rm erf} \left(a_{{5}}
\right )+{\rm erf} \left(a_{{7}}\right ) \right )
}
{
{\rm erf} \left(a_{{3}}\right )+{\rm erf} \left(a_{{4}}\right )
}
\quad .
\label{momento2lognormaltruncated}
\end{equation}
The two parameters $x_l$ and $x_u$ are the minimal and maximal elements of the sample.
The two parameters $m$ and $\sigma$ can be found
 through the MME,
 {\it first} method
\begin{equation}
E_T (X  ;m,\sigma,x_l,x_u)   = \bar{x}_1 \quad ;  \quad        
E_T (X^2;m,\sigma,x_l,x_u)   = \bar{x}_2  \quad .
\end{equation}
The above system  consists in
two non-linear functions in two variables and can therefore
be solved using the  Powell hybrid method,
see subroutine FORTRAN SNSQE in \cite{Kahaner1989}.
The {\it second}  method implements the MLE in order
to find $m$ and $\sigma$,
see Appendix \ref{appendixa}.

\section{Application to the stars}
This section reviews some useful statistical 
parameters, such as 
the merit function $\chi^2$,
the Akaike information criterion,
and
the Kolmogorov--Smirnov test.
The four samples of stars which test the truncated lognormal
distribution are introduced.

\subsection{The adopted statistics}
The merit function $\chi^2$
is computed
according to the formula
\begin{equation}
\chi^2 = \sum_{i=1}^n \frac { (T_i - O_i)^2} {T_i},
\label{chisquare}
\end {equation}
where $n  $   is the number of bins,
      $T_i$   is the theoretical value,
and   $O_i$   is the experimental value represented
by the frequencies.
The theoretical  frequency distribution is given by
\begin{equation}
 T_i  = N {\Delta x_i } p(x) \quad,
\label{frequenciesteo}
\end{equation}
where $N$ is the number of elements of the sample,
      $\Delta x_i $ is the magnitude of the size interval,
and   $p(x)$ is the PDF  under examination.
The size of the bins, $\Delta x_i $,
is equal for each bin
in the the case of linear histograms,
but different for each bin when
logarithmic  histograms  are considered.

A reduced  merit function $\chi_{red}^2$
is  evaluated  by
\begin{equation}
\chi_{red}^2 = \chi^2/NF
\quad,
\label{chisquarereduced}
\end{equation}
where $NF=n-k$ is the number of degrees  of freedom,
$n$ is the number of bins,
and $k$ is the number of parameters.
The goodness  of the fit can be expressed by
the probability $Q$, see  equation 15.2.12  in \cite{press},
which involves the degrees of freedom
and $\chi^2$.
According to  \cite{press} p.~658, the
fit ``may be acceptable'' if  $Q>0.001$. 

The Akaike information criterion
(AIC), see \cite{Akaike1974},
is defined by
\begin{equation}
AIC  = 2k - 2  ln(L)
\quad,
\end {equation}
where $L$ is
the likelihood  function  and $k$  the number of  free parameters
in the model.
We assume  a Gaussian distribution for  the errors
and  the likelihood  function
can be derived  from the $\chi^2$ statistic
$L \propto \exp (- \frac{\chi^2}{2} ) $
where  $\chi^2$ has been computed by
eqn.~(\ref{chisquare}),
see~\cite{Liddle2004}, \cite{Godlowski2005}.
Now the AIC becomes
\begin{equation}
AIC  = 2k + \chi^2
\quad.
\label{AIC}
\end {equation}

The Kolmogorov--Smirnov test (K--S),
see \cite{Kolmogoroff1941,Smirnov1948,Massey1951},
does not  require binning the data.
The K--S test,
as implemented by the FORTRAN subroutine KSONE in \cite{press},
finds
the maximum  distance, $D$, between the theoretical
and the astronomical  DF
as well the  significance  level  $P_{KS}$,
see formulas  14.3.5 and 14.3.9  in \cite{press};
if  $ P_{KS} \geq 0.1 $,
the goodness of the fit is believable.

\subsection{The selected sample of stars}

\label{applications}
\begin{figure*}
\begin{center}
\includegraphics[width=10cm]{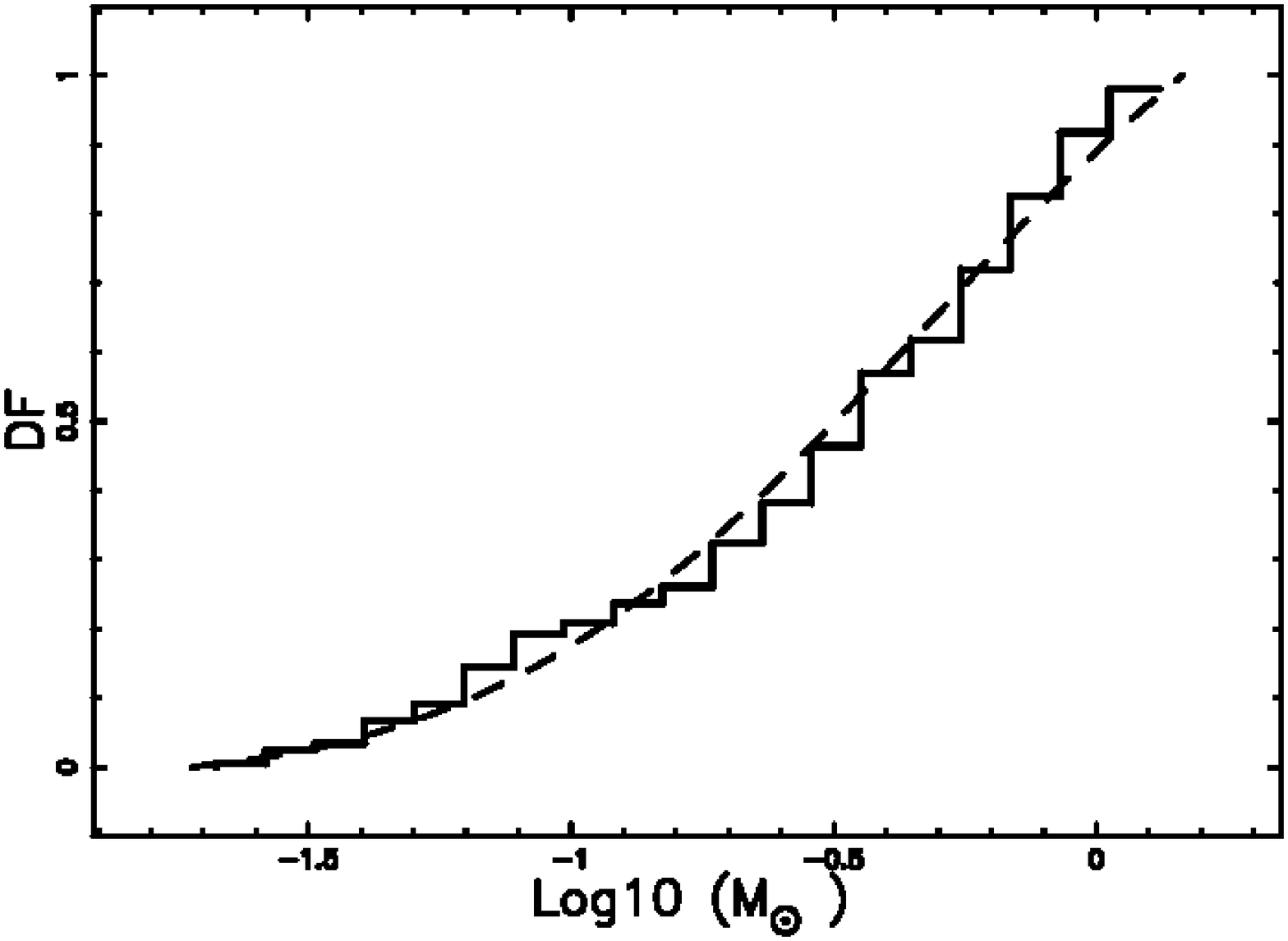}
\end{center}
\caption
{
Empirical DF   of  mass distribution
for    NGC 6611 cluster data 
(207 stars + BDs)
when the number of bins, $n$, is 20
(steps with full line)
with a superposition of the truncated lognormal     
DF  (dashed line).
Theoretical parameters as in Table \ref{chi2valuesngc6611}, MLE method.
The  horizontal axis  has a  logarithmic scale.
}
\label{lognormal_tronc_df_ngc6611}
\end{figure*}

\begin{figure*}
\begin{center}
\includegraphics[width=10cm]{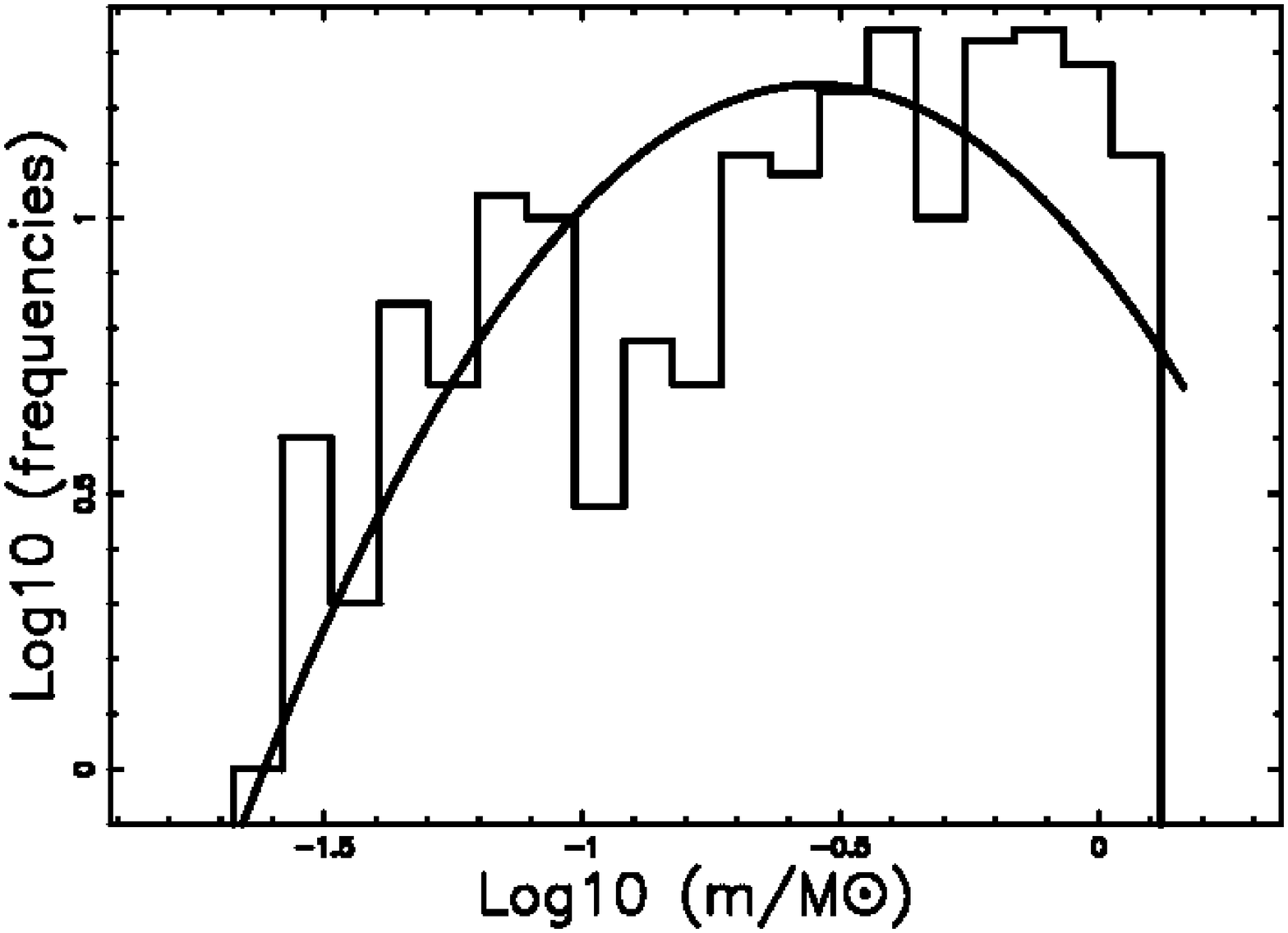}
\end{center}
\caption
{
Frequencies of    mass distribution
for    NGC 6611 cluster data 
(steps with full line)
with a superposition of the truncated lognormal     
PDF (full line).
Parameters as in Figure \ref{lognormal_tronc_df_ngc6611}.
The  vertical and horizontal axes  have  a  logarithmic   scale.
}
\label{lognormal_tronc_pdf_ngc6611}
\end{figure*}

The test samples are selected from the
Centre de Donn{\'e}es astronomiques de Strasbourg (CDS)  
in order to ensure that the test can be easily reproduced,
the name of the catalog is reported.
The first  test  is performed
on   the low-mass IMF
in the young cluster NGC 6611,
see  \cite{Oliveira2009} and CDS catalog J/MNRAS/392/1034.
This massive cluster has an age of 2--3 Myr and contains
masses from 
 $1.5  {M}_{\sun}~>~ {M} \geq  0.02  {M}_{\sun}$.
Therefore the brown dwarfs (BD)  region,
$\approx \, 0.2\,\sunmass$ is covered.
Table
\ref{chi2valuesngc6611}  shows the values of
$\chi_{red}^2$,
the
AIC,  the probability $Q$,
of  the  fits and the
two results of the K--S test:
the maximum  distance, $D$, between the theoretical
and the astronomical  DF
as well the  significance  level  $P_{KS}$.
Figure  
\ref{lognormal_tronc_df_ngc6611}  
shows the fit
with  the  truncated lognormal DF
for   NGC 6611, and Figure  
\ref{lognormal_tronc_pdf_ngc6611}
the truncated lognormal PDF.
\begin{table}[ht!]
\caption
{
Statistical parameters of
NGC 6611  (207 stars + BDs).
The  number of  linear   bins, $n$, is 20.
}
\label{chi2valuesngc6611}
\begin{center}
\resizebox{12cm}{!}
{
\begin{tabular}{|c|c|c|c|c|c|c|c|}
\hline
PDF       & Method &  parameters  &  AIC  & $\chi_{red}^2$
& $Q$  &  D &   $P_{KS}$  \\
\hline
lognormal &  MLE & $\sigma$=1.029, $m$=0.284   &  71.24&
3.73    & $1.3\,10^{-7}$  &  0.09366 &  0.04959 \\
\hline
lognormal &  MME & $\sigma$=0.676, $m$=0.339   &  107.46&
5.74   & $5.1\,10^{-14}$  &   0.172 &  $6.7\,10^{-6}$ \\
\hline
truncated lognormal & MLE & $\sigma$=1.499, $m$=0.478, $x_l$=0.0189, $x_u$=1.46   &
50.96   & 2.68      & $2.8\,10^{-4}$      &  0.0654 &  0.372
\\
\hline
truncated lognormal & MME & $\sigma$=0.977, $m$=0.361, $x_l$=0.0189, $x_u$=1.46   &
71.30   & 3.95      & $1.43\,10^{-7}$      &  0.117 &  0.005\\
\hline
\end{tabular}
}
\end{center}
\end{table}


\begin{figure*}
\begin{center}
\includegraphics[width=10cm]{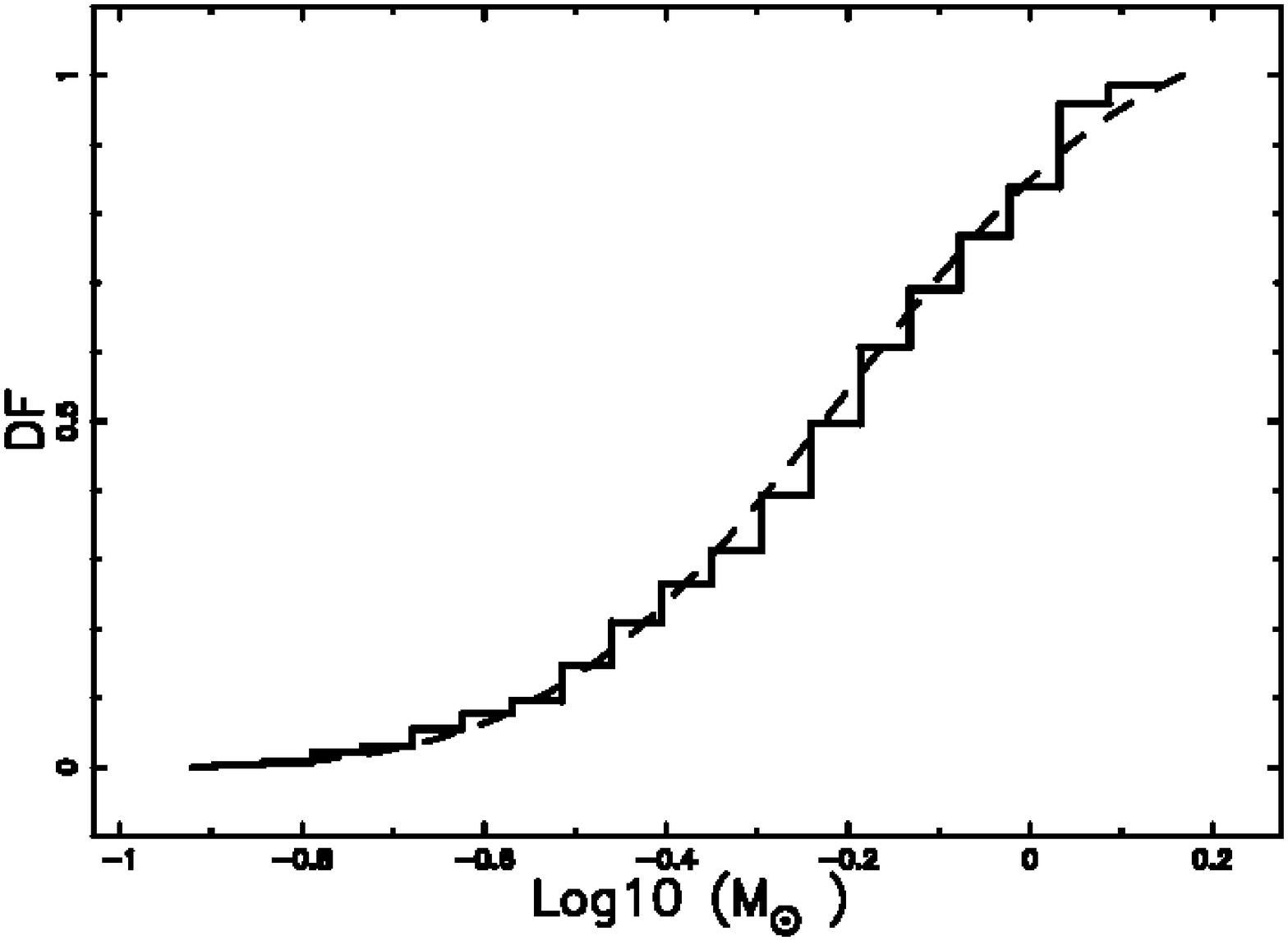}
\end{center}
\caption
{
Empirical DF   of  mass distribution
for   NGC 2362 cluster data (273 stars + BDs)
when the number of bins, $n$, is 20
(steps with full line)
with a superposition of the truncated lognormal  DF (dashed line).
Theoretical parameters as in Table \ref{chi2valuesngc2362}, MLE method.
The  horizontal axis  has a  logarithmic scale.
}
\label{lognormal_tronc_df_ngc2362}
\end{figure*}

\begin{figure*}
\begin{center}
\includegraphics[width=10cm]{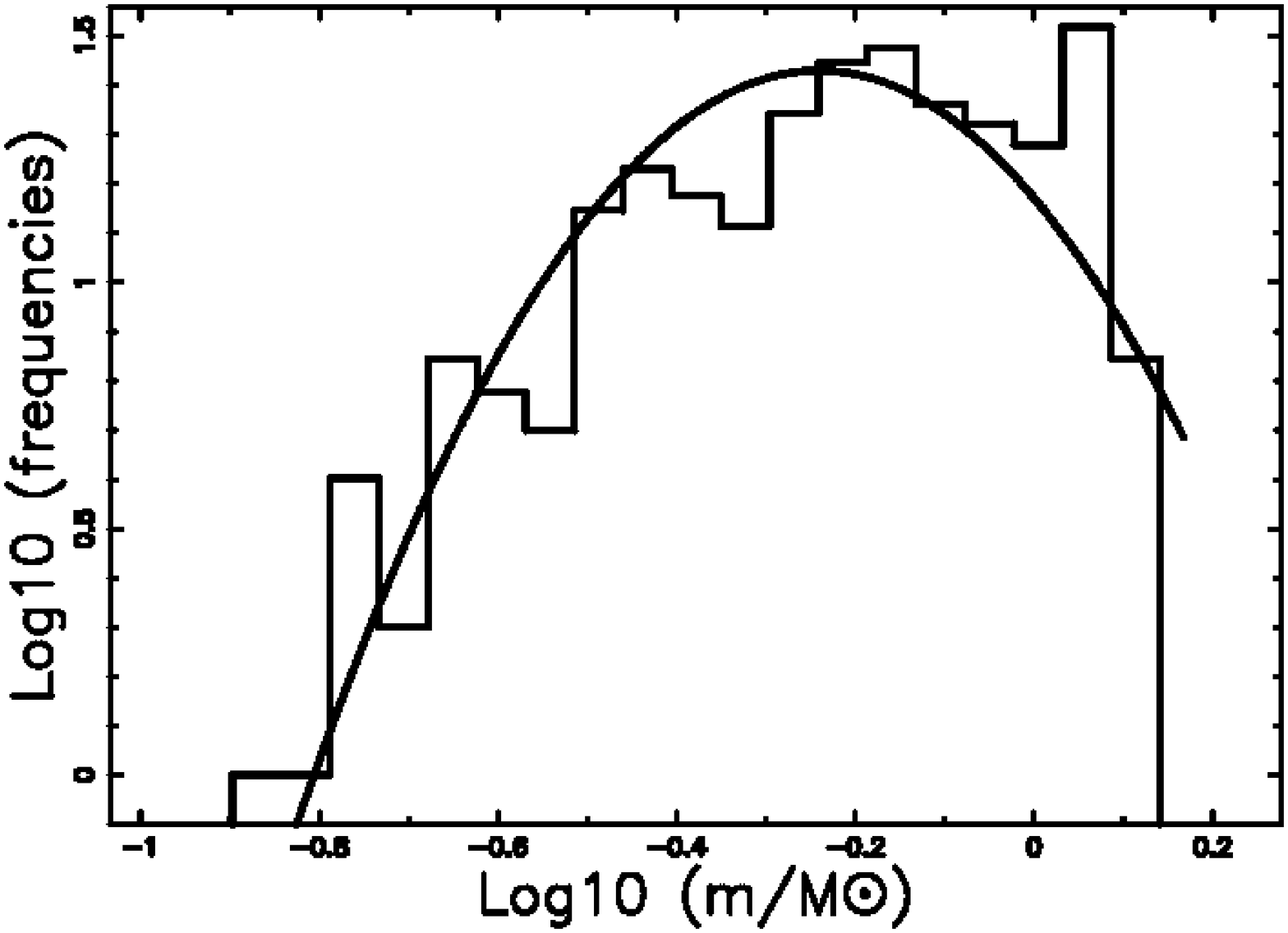}
\end{center}
\caption
{
Frequencies of   mass distribution
for    NGC 2362 cluster data 
(steps with full line)
with a superposition of the truncated lognormal     
PDF (full line).
Parameters as in Figure \ref{lognormal_tronc_df_ngc2362}.
The  vertical and horizontal axes  have  a  logarithmic   scale.
}
\label{lognormal_tronc_pdf_ngc2362}
\end{figure*}

The second  test  is performed
on  NGC 2362  where
the  271 stars
have a range
 $1.47  {M}_{\sun}~>~ {M} \geq  0.11  {M}_{\sun}$,
see  \cite{Irwin2008} and CDS catalog J/MNRAS/384/675/table1.
This is a  very young open cluster
with an estimated age of 3--9 Myr. 
Table
\ref{chi2valuesngc2362}
reports the statistical parameters,
\begin{table}[ht!]
\caption
{
Statistical parameters of
NGC 2362  (272 stars).
The  number of  linear   bins, $n$, is 20.
}
\label{chi2valuesngc2362}
\begin{center}
\resizebox{12cm}{!}
{
\begin{tabular}{|c|c|c|c|c|c|c|c|}
\hline
PDF       & Method &  parameters  &  AIC  & $\chi_{red}^2$
& $Q$  &  D &   $P_{KS}$  \\
\hline
lognormal &  MLE & $\sigma$=0.507, $m$=0.574 &  37.64&
1.86   & 0.013 & 0.072  & 0.105 \\
\hline
lognormal &  MME & $\sigma$ =0.428, $m$=0.588   & 51.66  & 2.648
   & $1.68\,10^{-4}$ & 0.0842   & 0.039   \\
\hline
truncated lognormal & MLE & $\sigma$=0.59, $m$= 0.625, $x_l$=0.119, $x_u$=1.47   &
 50.498 & 2.656     &$3.33 \,10^{-4}$   &0.047   &0.556
\\
\hline
truncated lognormal & MME & $\sigma$=0.521, $m$=0.612, $x_l$=0.119, $x_u$=1.47    &46.05
   & 2.37      &$1.4\,10^{-3}$       &0.048   & 0.525 \\
\hline
\end{tabular}
}
\end{center}
\end{table}
Figure  \ref{lognormal_tronc_df_ngc2362}  
shows the fit 
with  the  truncated lognormal DF
of  NGC 2362 and Figure  
\ref{lognormal_tronc_pdf_ngc2362} the fit with the
truncated lognormal PDF.

\begin{figure*}
\begin{center}
\includegraphics[width=10cm]{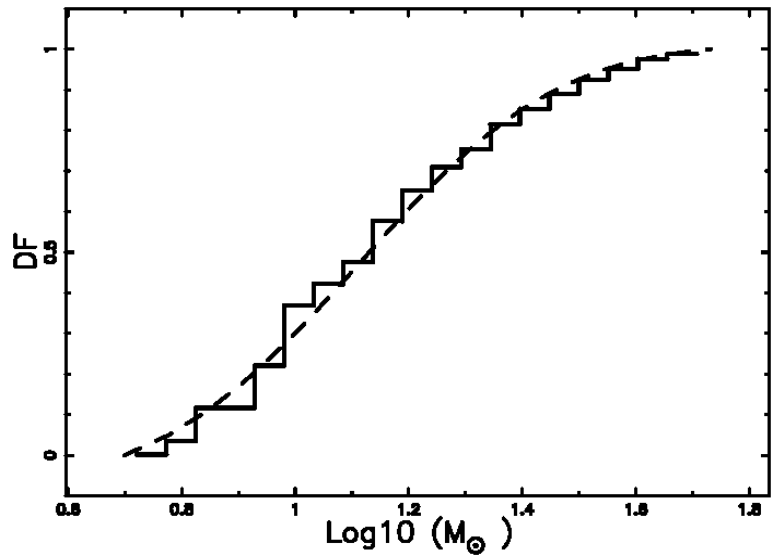}
\end{center}
\caption
{
Empirical DF   of  mass distribution
for  1563 stars in LMC
when the number of bins, $n$, is 20
(steps with full line)
with a superposition of the truncated lognormal 
distribution (dashed line).
Theoretical parameters as in Table \ref{chi2valuelmc}, MLE method.
The  horizontal axis  has a  logarithmic scale.
}
\label{lognormal_tronc_df_lmc}
\end{figure*}

\begin{figure*}
\begin{center}
\includegraphics[width=10cm]{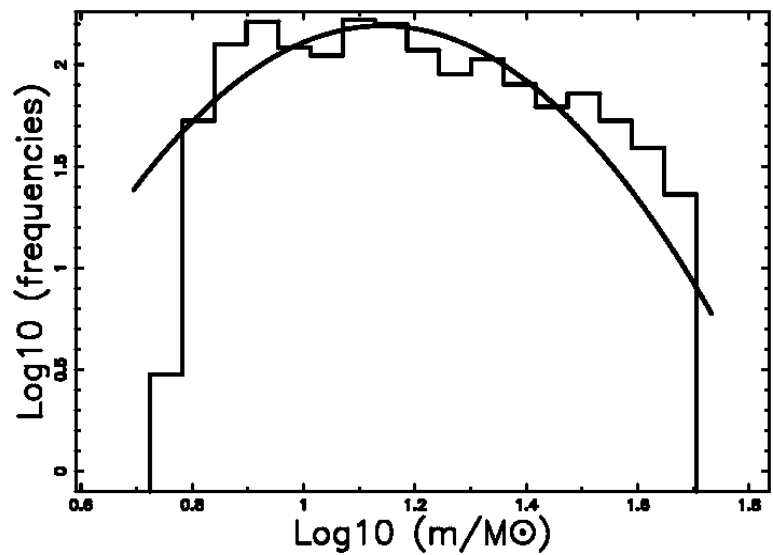}
\end{center}
\caption
{
Frequencies   of  mass distribution
for 1563 stars in LMC 
(steps with full line)
with a superposition of the truncated lognormal     
PDF (full line).
Parameters as in Figure \ref{lognormal_tronc_df_lmc}.
The  vertical and horizontal axes  have  a  logarithmic   scale.
}
\label{lognormal_tronc_pdf_lmc}
\end{figure*}

The third test  is performed
on  a
40$^{\prime}$
  circular field in the LMC made by 1563
 stars
in  the  range of masses, evaluated assuming an age of 4 Myr,
 $54 {M}_{\sun}~>~ {M} \geq  5  {M}_{\sun}$,
see  \cite{Hill1994}  and CDS catalog J/ApJ/425/122/table2.
Table
\ref{chi2valuelmc}
reports the statistical parameters.
\begin{table}[ht!]
\caption
{
Statistical parameters of a
circular field in the LMC
(1563 stars).
The  number of  linear   bins, $n$, is 20.
}
\label{chi2valuelmc}
\begin{center}
\resizebox{12cm}{!}
{
\begin{tabular}{|c|c|c|c|c|c|c|c|}
\hline
PDF       & Method &  parameters  &  AIC  & $\chi_{red}^2$
& $Q$  &  D &   $P_{KS}$  \\
\hline
lognormal &  MLE & $\sigma$=0.533, $m$= 13.84&  139.32& 7.51 & 5.07\,$10^{-20}$  & 0.0981 & 1.38\,$10^{-13}$  \\
\hline
lognormal &  MME & $\sigma$=0.554, $m$= 13.80&  122.07& 6.55& 9.6\,$10^{-17}$  & 0.0884 & 4.02\,$10^{-11}$  \\
\hline
truncated lognormal & MLE & $\sigma$=0.64, $m$= 12.9, $x_l$=5, $x_u$=54   &
 94.90 & 5.43     &$9.25 \,10^{-12}$   &0.073  & 8.24$\,10^{-8}$\\
\hline
truncated lognormal & MME & $\sigma$=0.7, $m$=12.36, $x_l$=5, $x_u$=54   &
  102.6& 5.91    &$3.51\,10^{-13}$   & 0.0895  & 2.25$\,10^{-11}$\\
\hline
\end{tabular}
}
\end{center}
\end{table}
Figures  \ref{lognormal_tronc_df_lmc}  
and     \ref{lognormal_tronc_pdf_lmc}
shows the fit
with  the  truncated lognormal DF and PDF respectively.

\begin{figure*}
\begin{center}
\includegraphics[width=10cm]{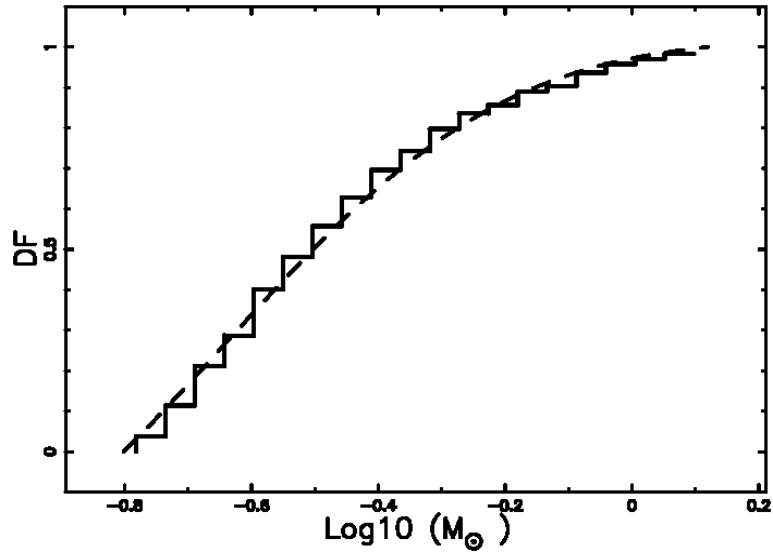}
\end{center}
\caption
{
Empirical DF   of  mass distribution
for  237  stars in $\gamma$ Velorum cluster 
when the number of bins, $n$, is 20
(steps with full line)
with a superposition of the truncated lognormal  DF (dashed line).
Theoretical parameters as in Table \ref{chi2valuegammavel}, MLE method.
The  horizontal axis  has a  logarithmic scale.
}
\label{lognormal_tronc_df_gamma_velorum}
\end{figure*}

\begin{figure*}
\begin{center}
\includegraphics[width=10cm]{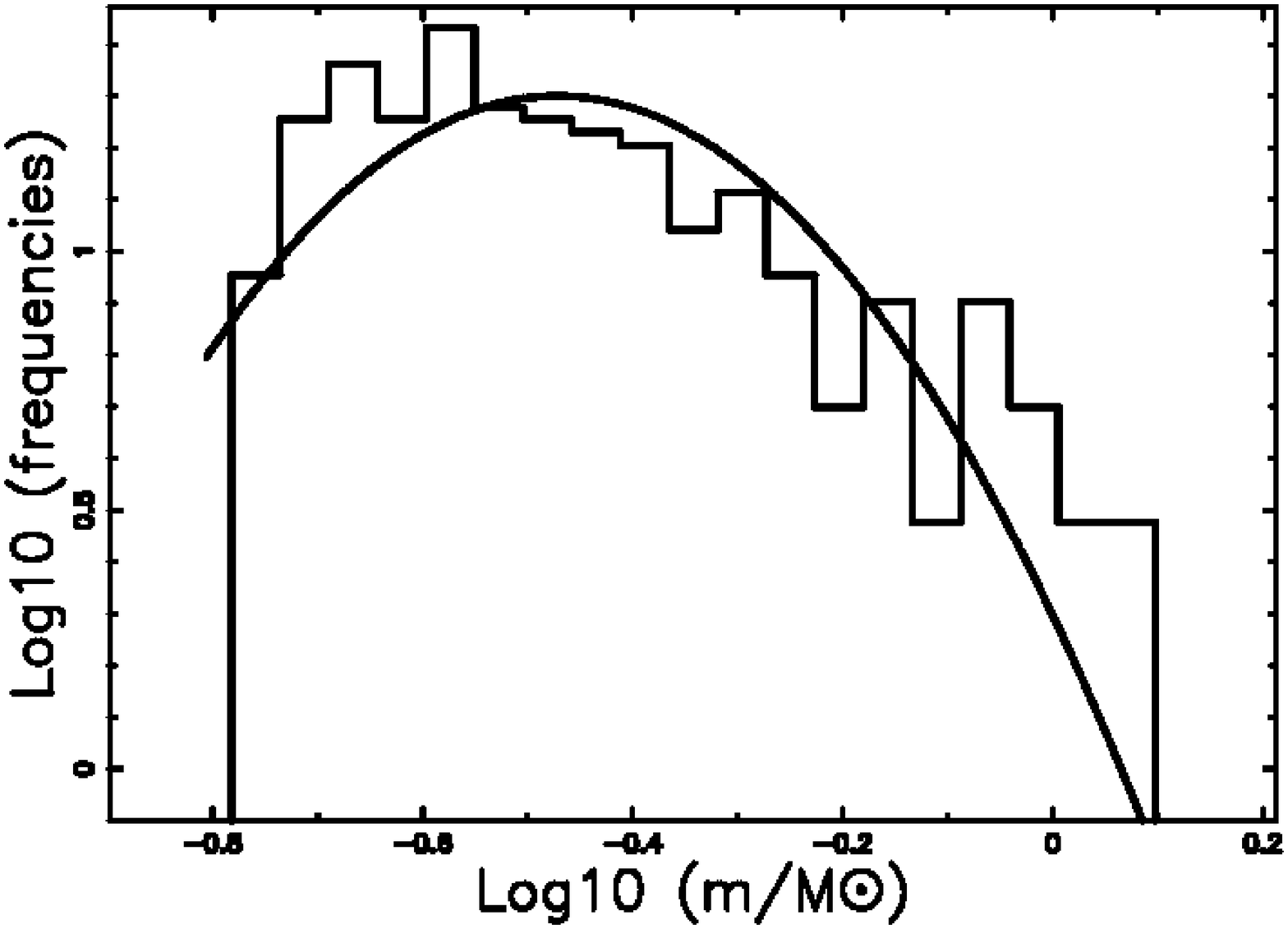}
\end{center}
\caption
{
Frequencies  of  mass distribution
for 237  stars in $\gamma$ Velorum cluster 
(steps with full line)
with a superposition of the truncated lognormal     
PDF (full line).
Parameters as in Figure \ref{lognormal_tronc_df_gamma_velorum}.
The  vertical and horizontal axes  have  a  logarithmic   scale.
}
\label{lognormal_tronc_pdf_gamma_velorum}
\end{figure*}

The fourth test  is performed
on  $\gamma$ Velorum cluster   where
the  237 stars
have a range
 $1.31 {M}_{\sun}~>~ {M} \geq  0.15  {M}_{\sun}$,
see  \cite{Prisinzano2016} and 
CDS catalog J/A+A/589/A70/table5.
This cluster is consists of 5--10 Myr old premain
sequence stars. 
The statistical  parameters are reported in 
Table \ref{chi2valuegammavel},
Figures 
\ref{lognormal_tronc_df_gamma_velorum}
and 
\ref{lognormal_tronc_pdf_gamma_velorum}
report the truncated lognormal DF and PDF respectively.

\begin{table}[ht!]
\caption
{
Statistical parameters of $\gamma$ Velorum cluster
(237 stars).
The  number of  linear   bins, $n$, is 20.
}
\label{chi2valuegammavel}
\begin{center}
\resizebox{12cm}{!}
{
\begin{tabular}{|c|c|c|c|c|c|c|c|}
\hline
PDF       & Method &  parameters  &  AIC  & $\chi_{red}^2$
& $Q$  &  D &   $P_{KS}$  \\
\hline
lognormal &  MLE & $\sigma$=0.504, $m$= 0.337 & 55.13 & 2.84 & 5.08\,$10^{-5}$
& 0.0921 & 0.0334  \\
\hline
lognormal &  MME & $\sigma$=0.564, $m$= 0.331 & 52.47 & 2.69 & 1.2\,$10^{-4}$
& 0.099 & 0.017  \\
\hline
truncated~lognormal &  MLE & $\sigma$=0.805, $m$= 0.227 & 30.54 & 1.4 
& 0.126  & 0.052 & 0.509  \\
\hline
truncated~lognormal &  MME & $\sigma$=0.504, $m$= 0.337 & 38.1 & 2.38 
& 1.4 \,$10^{-3}$ & 0.131 & 4.8\,$10^{-3}$   \\
\hline
\end{tabular}
}
\end{center}
\end{table}

\section{Other new distributions}

\begin{figure*}
\begin{center}
\includegraphics[width=10cm]{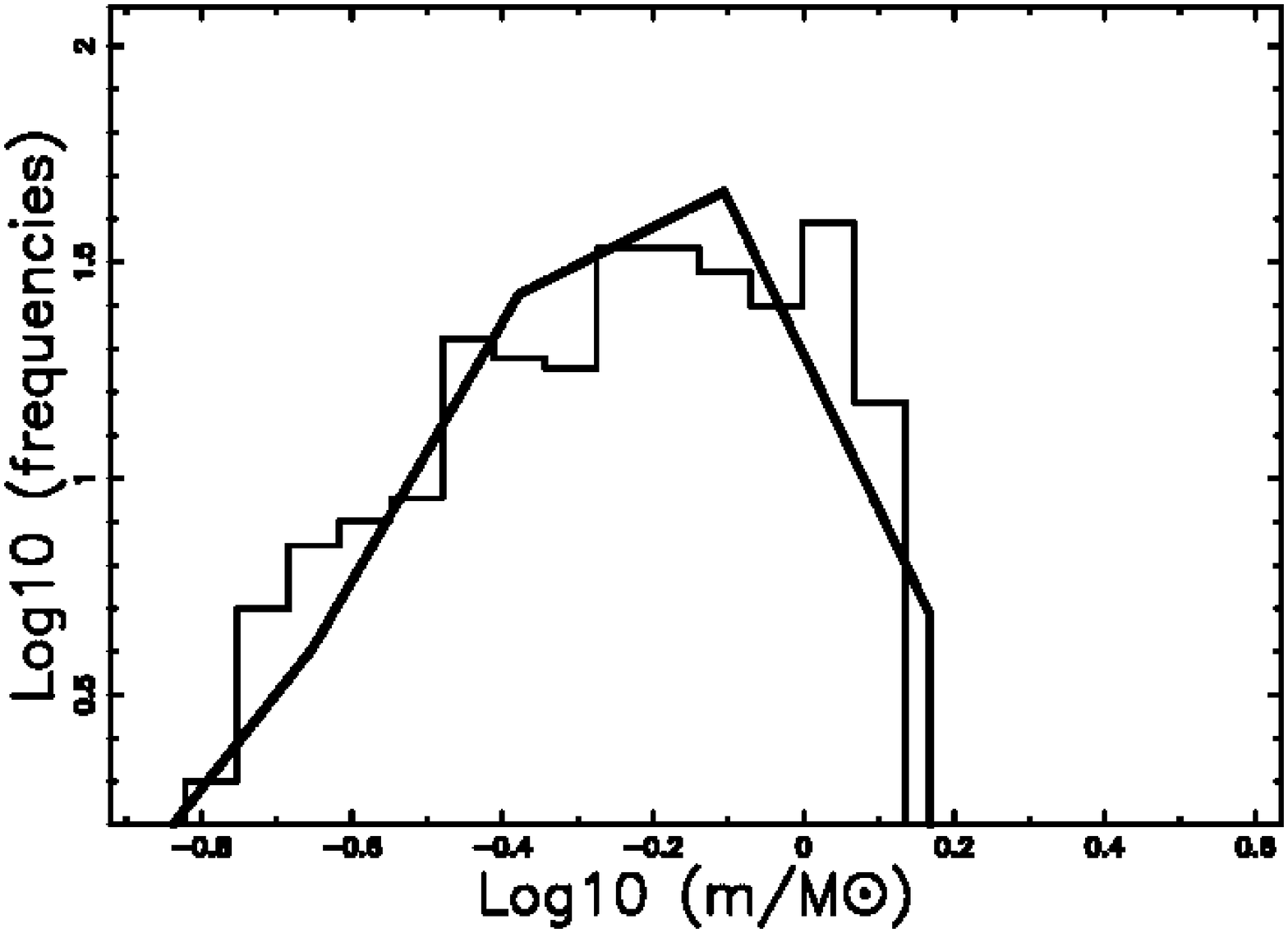}
\end{center}
\caption
{
Frequencies   of  mass distribution
for NGC 2362 cluster data (273 stars + BDs)
when the number of bins, $n$, is 16
(steps with full line)
with a superposition of the four piecewise inverse power law PDF (full line).
Theoretical parameters as in Table \ref{chi2valuesngc2362different}.
The  vertical and horizontal axes  have  a  logarithmic   scale.
}
\label{four_inverse_pdf_ngc2362}
\end{figure*}

As an initial  astronomical  reference, we display a 
piecewise broken inverse power law PDF,
see Figure \ref{four_inverse_pdf_ngc2362}
\label{others}

We now report  three recent PDFs.
The {\em first } is the  double Pareto lognormal
distribution which has
PDF
\begin{eqnarray}
f(x;\alpha,\beta,\mu,\sigma) =
\frac{1}{2}\,\alpha\,\beta\,( {{\rm e}^{\frac{1}{2}\,\alpha\,( \alpha\,{
\sigma}^{2}+2\,\mu-2\,\ln ( x )  ) }}{\it erfc}
( \frac{1}{2}\,{\frac {( \alpha\,{\sigma}^{2}+\mu-\ln ( x
 )  ) \sqrt {2}}{\sigma}} )
\nonumber \\
+{{\rm e}^{\frac{1}{2}\,\beta\,
( \beta\,{\sigma}^{2}-2\,\mu+2\,\ln ( x )  ) }
}{\it erfc}( \frac{1}{2}\,{\frac {( \beta\,{\sigma}^{2}-\mu+\ln
( x )  ) \sqrt {2}}{\sigma}} )  ) {x}^{-
1}( \alpha+\beta ) ^{-1},
\end{eqnarray}
where $\alpha$ and $\beta$
are the Pareto coefficients for the
upper and the lower tail,
respectively,
$\mu $ and $\sigma$  are
the lognormal body parameters, and
$erfc$  is
the complementary error function, see \cite{Reed2004}.
The mean ( for $\alpha >1$ ) can be expressed
as
\begin{equation}
E(\alpha,\beta,\mu,\sigma) =
\frac
{
\alpha\,\beta\,{{\rm e}^{\mu+\frac{1}{2}\,{\sigma}^{2}}}
}
{
\left( \alpha-1 \right)  \left( \beta+1 \right)
}
\quad.
\end{equation}

\begin{figure*}
\begin{center}
\includegraphics[width=10cm]{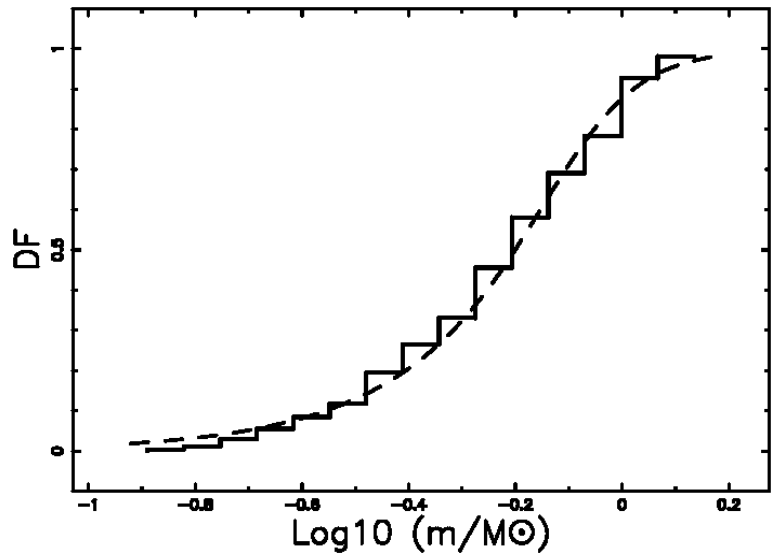}
\end{center}
\caption
{
Empirical  DF   of  mass distribution
for   NGC 2362 cluster data (273 stars + BDs)
when the number of bins, $n$, is 16
(steps with full line)
with a superposition of the double Pareto lognormal DF (full line).
Theoretical parameters as in Table \ref{chi2valuesngc2362different}.
The  horizontal axis  has a  logarithmic  scale.
}
\label{2log_pareto_df_ngc2362}
\end{figure*}

\begin{figure*}
\begin{center}
\includegraphics[width=10cm]{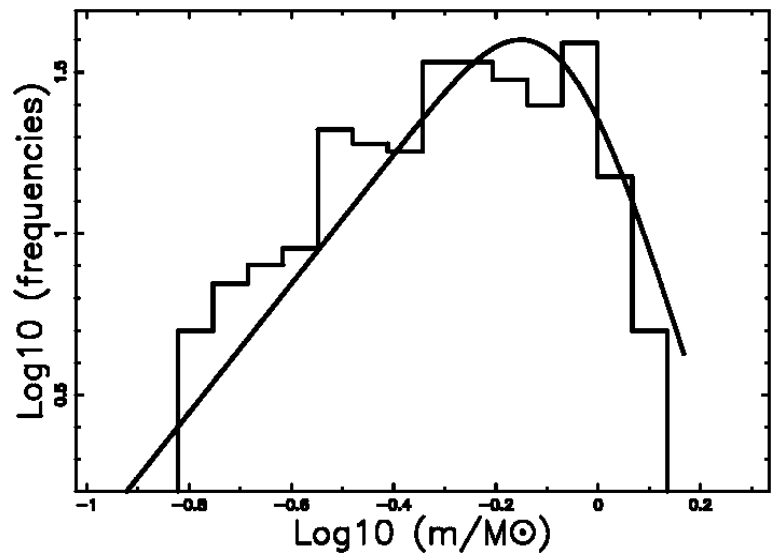}
\end{center}
\caption
{
Frequencies   of  mass distribution
for NGC 2362 cluster data (273 stars + BDs)
when the number of bins, $n$, is 16
(steps with full line)
with a superposition of the double Pareto lognormal PDF (full line).
Theoretical parameters as in Table \ref{chi2valuesngc2362different}.
The  vertical and horizontal axes  have  a  logarithmic   scale.
}
\label{2log_pareto_pdf_ngc2362}
\end{figure*}

This PDF exhibits a power law  behaviour in both tails
\begin{equation}
f(x) \sim k_1 \, x ^{-\alpha -1} (x \rightarrow \infty) \quad;
\quad
f(x) \sim k_2 \, x ^{\beta -1} (x \rightarrow 0 ) \quad,
\end{equation}
where $k_1$ and $k_2$ are two constants.
Figures 
\ref{2log_pareto_df_ngc2362}
and 
\ref{2log_pareto_pdf_ngc2362}
report the double Pareto lognormal DF and PDF respectively.

\begin{figure*}
\begin{center}
\includegraphics[width=10cm]{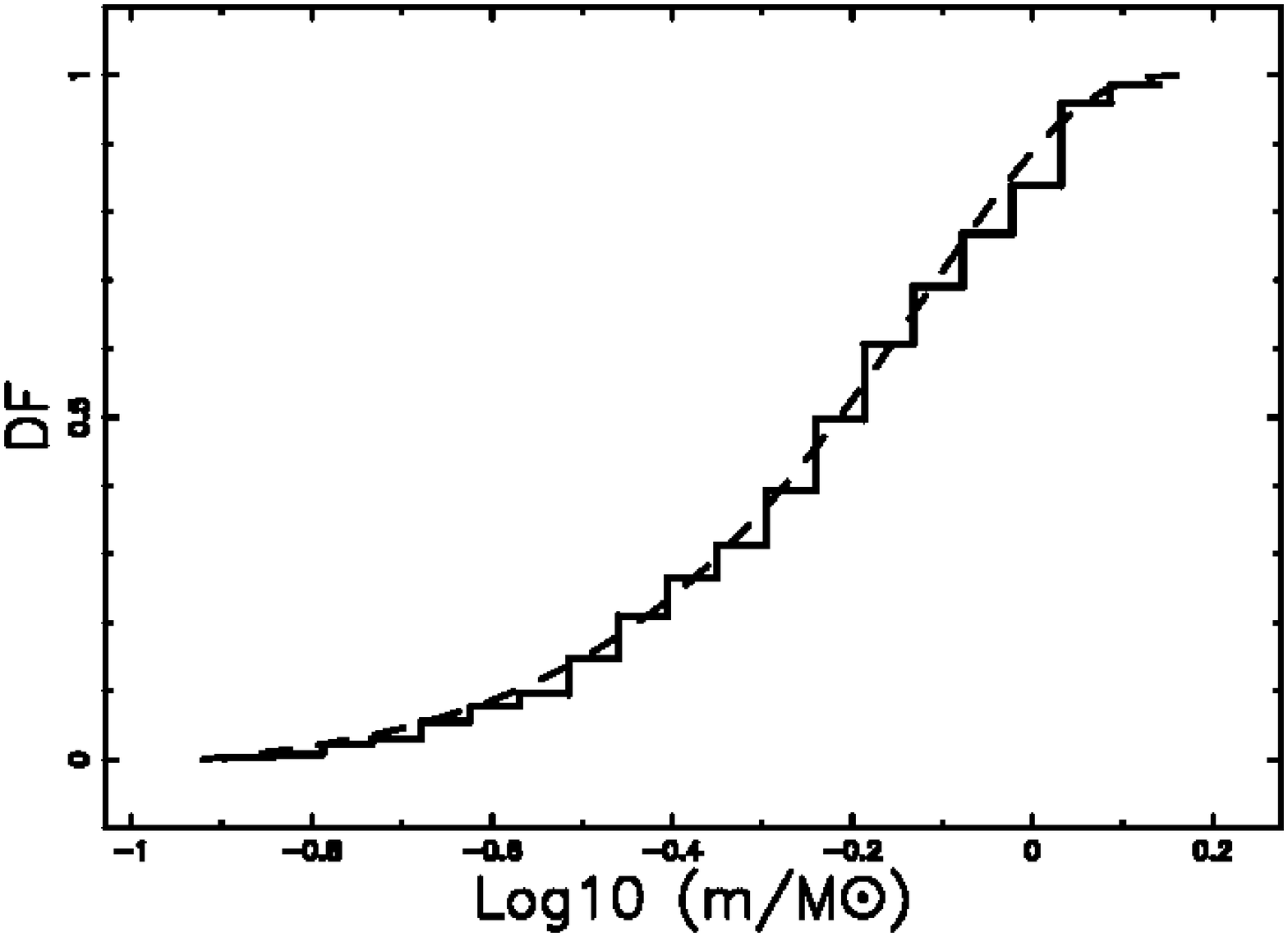}
\end{center}
\caption
{
Empirical  DF   of  mass distribution
for   NGC 2362 cluster data (273 stars + BDs)
when the number of bins, $n$, is 20
(steps at full line)
with a superposition of the left truncated beta DF (full line).
Theoretical parameters as in Table \ref{chi2valuesngc2362different}.
The  horizontal axis  has a  logarithmic   scale.
}
\label{beta_df_ngc2362}
\end{figure*}

\begin{figure*}
\begin{center}
\includegraphics[width=10cm]{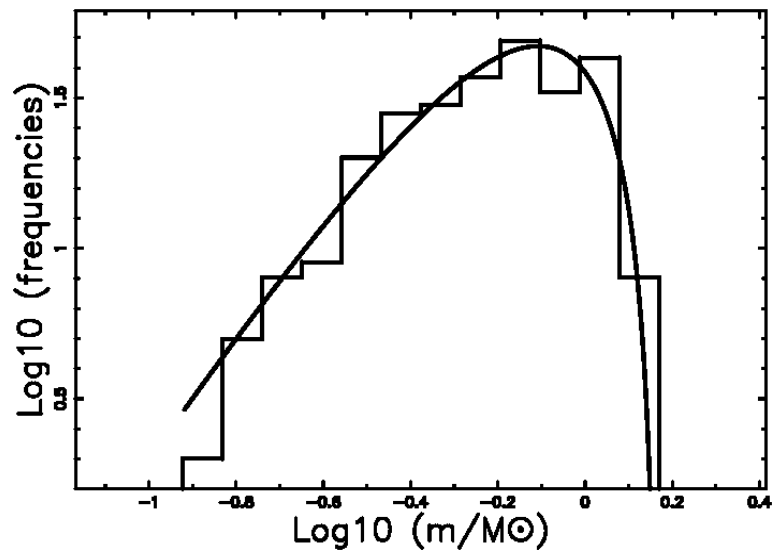}
\end{center}
\caption
{
Frequencies   for   mass distribution
in  NGC 2362 cluster  with
(full line steps) 
and left truncated beta PDF.
Parameters as in Figure \ref{beta_df_ngc2362}.
The  vertical and horizontal axes  have  a  logarithmic   scale.
}
\label{beta_pdflog_ngc2362}
\end{figure*}

The {\em second } is the left truncated beta with scale PDF
which is
\begin {equation}
f_T(x;a,b,\alpha,\beta) = K\,{x}^{\alpha-1} \left( b-x \right )
^{\beta-1},
\label{betatruncated}
\end{equation}
where the constant is
\begin{equation}
K=\frac{-\alpha\,\Gamma  \left( \alpha+\beta \right )}
{{b}^{\beta-1} H\,{a}^{ \alpha}\Gamma  \left( \alpha+\beta \right )
-{b}^{\beta-1+\alpha} \Gamma  \left( 1+\alpha \right ) \Gamma
\left( \beta \right )},
\end{equation}
and
\begin{equation}
 H={\mbox{$_2$F$_1$}(\alpha,-\beta+1;\,1+\alpha;\,{\frac
{a}{b}})} \quad ,
\end{equation}
where ${\2F1(a,b;\,c;\,z)}$ is the
regularized hypergeometric
function \cite{Abramowitz1965},
see \cite{Zaninetti2013a}.
Figure 
\ref{beta_df_ngc2362} 
reports the
DF and Figure 
\ref{beta_pdflog_ngc2362} 
the PDF.

The {\em third}  is
the truncated gamma  (TG) PDF  which is
\begin {equation}
f(x;b,c,x_l,x_u) =
k\; \left( {\frac {x}{b}} \right ) ^{c-1}{{\rm e}^{-{\frac {x}{b}}}}
\label{gammatruncated}
\end {equation}
where  the constant  $k$
is
\begin{equation}
k =
\frac{c}
{
b\Gamma  \left( 1+c,{\frac {x_{{l}}}{b}} \right ) -b\Gamma  \left( 1+c,
{\frac {x_{{u}}}{b}} \right ) +{{\rm e}^{-{\frac {x_{{u}}}{b}}}}{b}^{-c
+1}{x_{{u}}}^{c}-{{\rm e}^{-{\frac {x_{{l}}}{b}}}}{b}^{-c+1}{x_{{l}}}^
{c}
}
\quad  ,
\label{constant}
\end {equation}
where
\begin{equation}
\mathop{\Gamma\/}\nolimits\!\left(a,z\right )=\int_{z}^{\infty}t^{{a-1}}e^{{-t}%
}dt,
\end{equation}
is the  upper incomplete gamma function,
see \cite{Zaninetti2013e}.

\begin{figure*}
\begin{center}
\includegraphics[width=10cm]{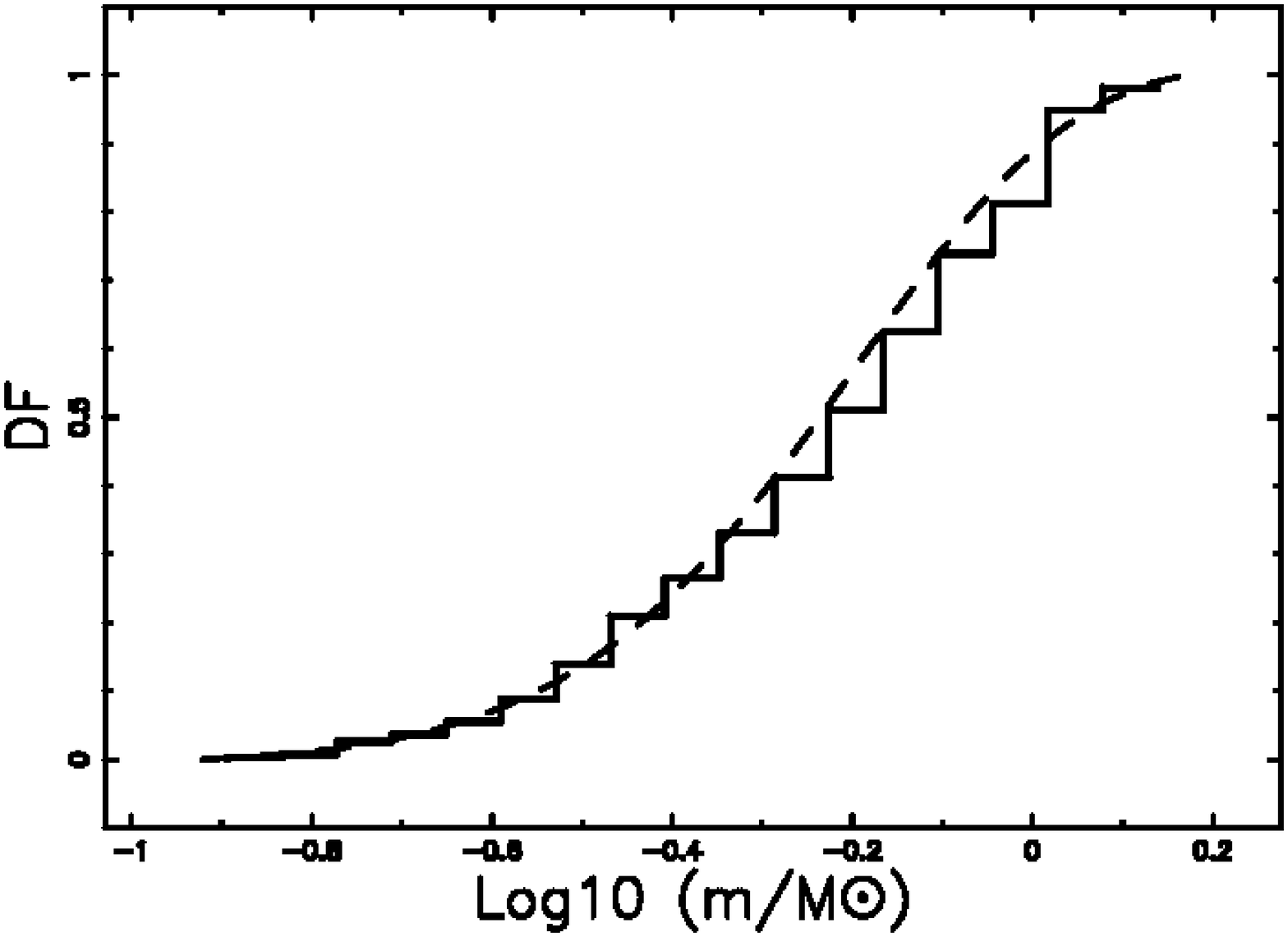}
\end{center}
\caption
{
Empirical  DF   of  mass distribution
for    NGC 2362 cluster data (273 stars + BDs)
when the number of bins, $n$, is 18
(steps at full line)
with a superposition of the truncated gamma DF (full line).
Theoretical parameters as in Table \ref{chi2valuesngc2362different}.
The  horizontal axis  has a  logarithmic   scale.
}
\label{gamma_tronc_df_ngc2362}
\end{figure*}

\begin{figure*}
\begin{center}
\includegraphics[width=10cm]{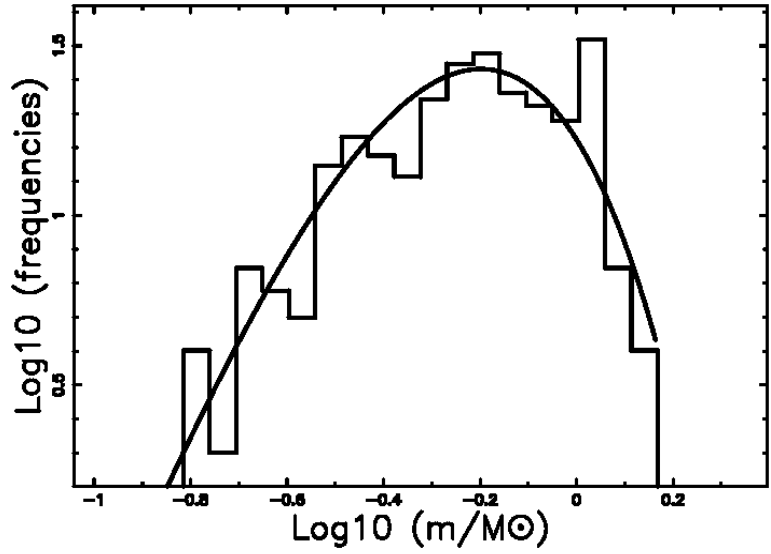}
\end{center}
\caption
{
Frequencies   of  mass distribution
for   NGC 2362 cluster data (273 stars + BDs)
when the number of bins, $n$, is 18
(steps at full line)
with a superposition of the truncated gamma PDF (full line).
Theoretical parameters as in Table \ref{chi2valuesngc2362different}.
The  vertical and horizontal axes  have  a  logarithmic   scale.
}
\label{gamma_tronc_pdf_ngc2362}
\end{figure*}
\begin{figure*}
\begin{center}
\includegraphics[width=10cm]{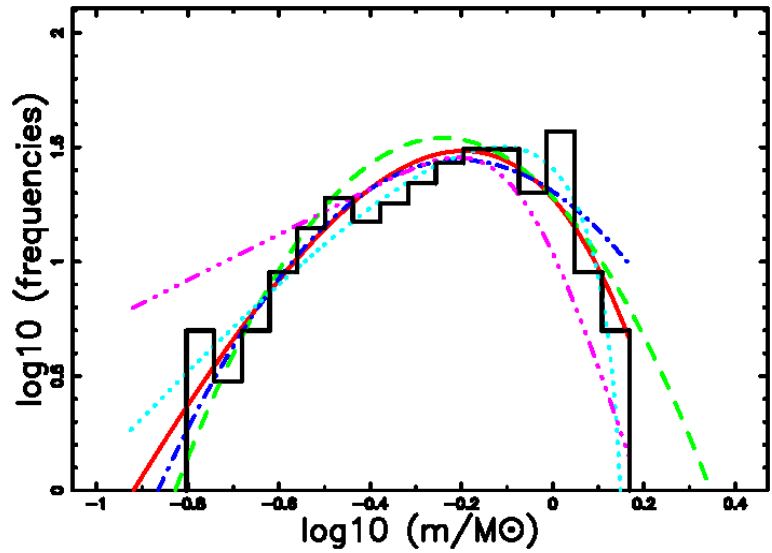}
\end{center}
\caption
{
Histogram (black step-diagram)        of  mass distribution
as  given by  NGC 2362 cluster data (273 stars +BDs )
with a superposition of 
the  truncated gamma PDF        (full red line),
the lognormal PDF               (dashed green line ),
the truncated lognormal     PDF (dot-dash-dot-dash blue line),
the truncated beta          PDF (dotted cyan line) 
and
the double Pareto lognormal PDF (dash-dot-dot-dot magenta line).
Vertical and horizontal axes have logarithmic scales.
}
\label{all_pdf_ngc2362}
\end{figure*}

Figure 
\ref{gamma_tronc_df_ngc2362} 
reports the  truncated gamma 
DF and  
Figure 
\ref{gamma_tronc_pdf_ngc2362} the truncated gamma PDF.
Table \ref{chi2valuesngc2362different}  reports
the parameters of these three new PDFs as well
as the parameters of the truncated lognormal
in the case of NGC 2362.
\begin{table}[ht!]
\caption
{
Statistical parameters of
NGC 2362  (272 stars) for different distributions
}
\label{chi2valuesngc2362different}
\begin{center}
\resizebox{12cm}{!}
{
\begin{tabular}{|c|c|c|c|}
\hline
PDF       &  parameters   &  D &   $P_{KS}$  \\
\hline
truncated lognormal  &
$\sigma$=0.59, $m$= 0.625, $x_l$=0.119, $x_u$=1.47
& 0.047  & 0.556
\\
\hline
truncated ~gamma & $b=0.159$, $c =4$, $x_l$=0.12, $x_u$=1.47
& 0.067  & 0.158
\\
\hline
double ~Pareto-lognormal                                 &
$\alpha$=5, $\beta$=2, $\sigma$=0.207, $\mu_{LN}$ =-0.25  &
0.05    & 0.471 \\
\hline
left~ truncated ~beta                           &
$a=0.12 $, $b=$1.47, $\alpha=2.18 $, $\beta$=2.93 &
  0.048   & 0.53   \\
\hline
four~inverse~power~law &
$m_1$ = 0.11,~$m_2=0.22$,~$m3=0.41$, $m_4=0.78$,~$m_5=1.47$
& 0.081 & 0.052    \\
~              & $\alpha_1=-1.2~,\alpha_2=-1.98,
\alpha_3=0.12,\alpha_4=4.57$ 
& & \\
\hline
\end{tabular}
}
\end{center}
\end{table}
 can be found 
Figure \ref{all_pdf_ngc2362}   displays 
all the PDFs here analysed.

\section{Conclusions}

The truncated  lognormal distribution gives better results,
i.e. higher $P_{KS}$, 
than the lognormal distribution,
see Tables
\ref{chi2valuesngc6611},
\ref{chi2valuesngc2362},
\ref{chi2valuelmc} 
and 
\ref{chi2valuegammavel}.
for the samples here considered.
The lower and upper boundaries in mass are connected with the
physical theories on the minimum and maximum mass
for the stars.
Fisher's conjecture (see \cite{Hald1999}) that statistical parameters
are better inferred through the maximum likelihood estimator (MLE) 
than through the matching of moments estimator (MME) is also tested:
in eight  cases out of eight,  the MLE produces better results,
see Tables
\ref{chi2valuesngc6611},
\ref{chi2valuesngc2362},
\ref{chi2valuelmc} 
and 
\ref{chi2valuegammavel}.
The comparison of the truncated lognormal DF with
other DFs  assigns the best results
to the truncated lognormal,
i.e. higher $P_{KS}$, 
even if
the difference from the double Pareto lognormal is small,
see Table \ref{chi2valuesngc2362different}.

The number of free parameters of the truncated lognormal 
PDF is {\it two} once the lower and upper
boundary are associated with the minimum and maximum 
mass of the considered sample,
see \ref{appendixa} for the MLE method.
In contrast, the number 
of parameters of the widely used 
four-piecewise broken inverse power law IMF 
is {\it seven}.

\appendix

\section{The parameters of the truncated lognormal}
\label{appendixa}

The parameters of the truncated lognormal distribution
can be obtained from empirical data by
the
maximum likelihood estimators  (MLE) and by the evaluation of the minimum and maximum elements
of the sample.
Consider a  sample  ${\mathcal X}=x_1, x_2, \dots , x_n$ and let
$x_{(1)} \geq x_{(2)} \geq \dots \geq x_{(n)}$ denote
their order statistics, so that
$x_{(1)}=\max(x_1, x_2, \dots, x_n)$, $x_{(n)}=\min(x_1, x_2, \dots, x_n)$.
The first two parameters $x_l$ and $x_u$
are
\begin{equation}
{x_l}=x_{(n)}, \qquad { x_u}=x_{(1)}
\quad  .
\label{eq:firstpar}
\end{equation}
The MLE is obtained by maximizing
\begin{equation}
\Lambda = \sum_i^n \ln(TL(x_i;m,\sigma,x_l,x_u)).
\end{equation}
The two derivatives $\frac{\partial \Lambda}{\partial m} =0$ and
$\frac{\partial \Lambda}{\partial \sigma}=0 $  generate two
non-linear equations in
 $m$ and $\sigma$ which can be solved numerically,
 we used FORTRAN subroutine SNSQE in \cite{Kahaner1989},
\begin{eqnarray}
\frac{\partial \Lambda}{\partial m}=
   ( {\rm erf}   (\frac{1}{2}\,{\frac {\sqrt {2}   ( \ln    ( x_{{
l}}   ) -\ln    ( m   )    ) }{\sigma}}  )-
{\rm erf}   (\frac{1}{2}\,{\frac {\sqrt {2}   ( \ln    ( x_{{u}}
   ) -\ln    ( m   )    ) }{\sigma}}  )   )
  \nonumber \\
   ( n\sqrt {2}\sigma\,{{\rm e}^{-\frac{1}{2}\,{\frac {   ( \ln    (
x_{{l}}   ) -\ln    ( m   )    ) ^{2}}{{\sigma}^{2}}}}}
-n\sqrt {2}\sigma\,{{\rm e}^{-\frac{1}{2}\,{\frac {   ( \ln    ( x_{{u}}
   ) -\ln    ( m   )    ) ^{2}}{{\sigma}^{2}}}}}
\nonumber \\
-\sqrt
{\pi}   ( {\rm erf}   (\frac{1}{2}\,{\frac {\sqrt {2}   ( \ln
   ( x_{{l}}   ) -\ln    ( m   )    ) }{\sigma}}
  )
 \nonumber \\
  -{\rm erf}   (\frac{1}{2}\,{\frac {\sqrt {2}   ( \ln    ( x_{{
u}}   ) -\ln    ( m   )    ) }{\sigma}}  )   )
   ( n\ln    ( m   ) -\sum _{i=1}^{n}\ln    ( x_{{i}}
   )    )    ) =0
\quad ,
\end{eqnarray}
and
\begin{equation}
\frac{\partial \Lambda}{\partial \sigma}= \frac{N}{D} =0,
\end{equation}
where
\begin{eqnarray}
N = \ln  \left( x_{{u}} \right) \sqrt {2}{{\rm e}^{-\frac{1}{2}\,{\frac {   (
\ln    ( x_{{u}}   ) -\ln    ( m   )    ) ^{2}}{{
\sigma}^{2}}}}}n\sigma-\ln    ( x_{{l}}   ) \sqrt {2}{{\rm e}^{
-\frac{1}{2}\,{\frac {   ( \ln    ( x_{{l}}   )
-\ln    ( m
   )    ) ^{2}}{{\sigma}^{2}}}}}n\sigma
\nonumber \\
   +\sqrt {2}{{\rm e}^{-1/
2\,{\frac {   ( \ln    ( x_{{l}}   ) -\ln    ( m   )
   ) ^{2}}{{\sigma}^{2}}}}}\ln    ( m   ) n\sigma-\sqrt {2}
{{\rm e}^{-\frac{1}{2}\,{\frac {   ( \ln    ( x_{{u}}   ) -\ln
   ( m   )    ) ^{2}}{{\sigma}^{2}}}}}\ln    ( m
   ) n\sigma
\nonumber \\
   +n   ( \ln    ( m   )    ) ^{2}\sqrt {
\pi}{\rm erf}   (\frac{1}{2}\,{\frac {\sqrt {2}   ( \ln    ( x_{{u}}
   ) -\ln    ( m   )    ) }{\sigma}}  )
\nonumber  \\
-n{\sigma}^{
2}\sqrt {\pi}{\rm erf}   (\frac{1}{2}\,{\frac {\sqrt {2}   ( \ln
   ( x_{{u}}   ) -\ln    ( m   )    ) }{\sigma}}
  )
\nonumber \\
  -n   ( \ln    ( m   )    ) ^{2}\sqrt {\pi}
{\rm erf}   (\frac{1}{2}\,{\frac {\sqrt {2}   ( \ln    ( x_{{l}}
   ) -\ln    ( m   )    ) }{\sigma}}  )
\nonumber \\
+n{\sigma}^{
2}\sqrt {\pi}{\rm erf}   (\frac{1}{2}\,{\frac {\sqrt {2}   ( \ln
   ( x_{{l}}   ) -\ln    ( m   )    ) }{\sigma}}
  )
\nonumber \\
  +\sum _{i=1}^{n}\ln    ( x_{{i}}   )    ( \ln
   ( x_{{i}}   ) -2\,\ln    ( m   )    ) \sqrt {\pi}
{\rm erf}   (\frac{1}{2}\,{\frac {\sqrt {2}   ( \ln    ( x_{{u}}
   ) -\ln    ( m   )    ) }{\sigma}}  )
\nonumber \\
   -\sum _{i=1
}^{n}\ln    ( x_{{i}}   )    ( \ln    ( x_{{i}}   ) -
2\,\ln    ( m   )    ) \sqrt {\pi}{\rm erf}   (\frac{1}{2}\,{
\frac {\sqrt {2}   ( \ln    ( x_{{l}}   ) -\ln    ( m
   )    ) }{\sigma}}  )
\quad ,
\end{eqnarray}

\begin{eqnarray}
D=\sqrt {\pi}   \Bigg ( -{\rm erf}   
 \bigg (\frac{1}{2}\,{\frac {\sqrt {2}   ( \ln
   ( x_{{l}}   ) -\ln    ( m   )    ) }{\sigma}}
 \bigg )  
\nonumber \\ 
+{\rm erf}  \bigg  (\frac{1}{2}\,{\frac {\sqrt {2}   ( \ln    ( x_{{
u}}   ) -\ln    ( m   )    ) }{\sigma}} \bigg )  \Bigg )
{\sigma}^{3} \quad .
\end{eqnarray}


\end{document}